\documentclass[aps,preprintnumbers,twocolumn,amsmath,nofootinbib,amssymb]{revtex4}
\usepackage{graphicx,color,dcolumn,booktabs,bm}
\usepackage{longtable,lscape}
\usepackage{txfonts}
\usepackage{overpic}
\usepackage{amssymb}
\usepackage{epstopdf}
\usepackage{indentfirst}
\usepackage{feynmf}   
\usepackage{slashed}  
\usepackage{cases}
\usepackage{color}
\usepackage{float}
\usepackage{multirow}
\usepackage{epsfig,amssymb,amsmath,amsfonts,amsbsy,mathrsfs}
\usepackage{gensymb}

\graphicspath{{Figures/}} %

\usepackage{hyperref}

\makeatletter
\@addtoreset{equation}{section}
\makeatother

\allowdisplaybreaks

\begin{document}

\title{Role of the low-lying nucleon resonances in the $p\bar{p} \to \psi \eta$ reaction}
\author{Qin-Song Zhou$^{1,2}$}\email{zhouqs13@lzu.edu.cn}
\author{Jun-Zhang Wang$^{1,2}$}\email{wangjzh2012@lzu.edu.cn}
\author{Ju-Jun Xie$^{3,2,4,5}$\footnote{Corresponding author}}\email{xiejujun@impcas.ac.cn}
\author{Xiang Liu$^{1,2}$\footnote{Corresponding author}}\email{xiangliu@lzu.edu.cn}
\affiliation{
$^1$School of Physical Science and Technology, Lanzhou University, Lanzhou 730000, China}
\affiliation{
$^2$Research Center for Hadron and CSR Physics, Lanzhou University
and Institute of Modern Physics of CAS, Lanzhou 730000, China}
\affiliation{$^3$Institute of Modern
Physics, Chinese Academy of Sciences, Lanzhou 730000, China}
\affiliation{$^4$School of Nuclear Science and Technology, University of Chinese Academy of Sciences, Beijing
100049, China} \affiliation{$^5$School of Physics and Microelectronics, Zhengzhou
University, Zhengzhou, Henan 450001, China}

\date{\today}
\begin{abstract}
Within the effective Lagrangian approach, we study the $p\bar{p} \rightarrow \psi\eta$ [$\psi\equiv\psi(3686)$, $J/\psi$] reaction at the low energy where the contributions from nucleon pole and low-lying nucleon resonances, $N(1520)$, $N(1535)$ and $N(1650)$ are considered. All the model parameters are determined with the help of current experimental data on the decay of $\psi \to p \bar{p} \eta$. Within the model parameters, the total and differential cross sections of the $p\bar{p} \rightarrow \psi\eta$ reaction are predicted. We show that the relative phases between different amplitudes of different nucleon resonance will change significantly the angular distributions of the $\bar{p} p \to \psi \eta$ reaction. Therefore, we conclude that these reactions are suitable to study experimentally the properties of the low-lying nucleon resonance and the reaction mechanisms. We hope that these theoretical calculations can be tested by future experiments.
\end{abstract}
\maketitle

\section{Introduction}\label{Introduction}

The charmonium, $H_c$, production in the $p\bar{p} \to H_c X$ (the $X$ is a light meson) reaction is an interesting tool to gain a deeper understanding of the strong interaction and also of the nature of the hadrons~\cite{Lundborg:2005am}. There is a forthcoming experimental effort, the Anti-Proton Annihilations at Darmstadt ($\rm{\bar{P}}$ANDA), dedicated to this reaction~\cite{Lutz:2009ff}. On the theoretical side, there exist several previous studies of this reaction. In Ref. \cite{Gaillard:1982zm}, Gaillard and Maiani first estimated differential cross sections for the process of $p\bar{p} \to \psi \pi^{0}$ in the soft pion limit. In Refs.~\cite{Barnes:2006ck,Barnes:2007ub,Barnes:2010yb}, it was pointed out that, with these $N^*N\psi$ couplings extracted form the corresponding $\psi \rightarrow \bar{p}N^*$ decays, the contributions of intermediated $N^*$ resonances and the nucleon pole to the process $p\bar{p}\rightarrow \psi X$ can be investigated. Then, Lin, Xu, and Liu~\cite{Lin:2012ru} considered the contributions of the intermediate nucleon pole and the effect of form factors (FFs) on charmonium production in the low-energy $p\bar{p}$ interaction at $\bar{\mbox{P}}$ANDA. It was shown that the effect of the FFs is significant. In Ref. \cite{Pire:2013jva}, Pire {\it et al.} studied the associated production of a $J/\psi$ and a $\pi$ through antiproton-nucleon annihilation in the framework of QCD collinear factorization.

Since the experimental data on the $\psi \to p \bar{p} X$ decays become rich, we can consider the contributions from nucleon resonances in the $p\bar{p}\rightarrow \psi X$ reaction where parameters can be fixed through the process of $\psi\rightarrow p\bar{p} X$. Indeed, in Refs.~\cite{Wiele:2013vla,Xu:2015qqa,Wang:2017sxq}, the authors have calculated the cross sections of the processes $p\bar{p}\rightarrow J/\psi \pi^{0}$, $p\bar{p}\rightarrow \psi(3770) \pi^{0}$ and $p\bar{p}\rightarrow Y(4220) \pi^{0}$, respectively, where the contributions from the intermediate nucleon resonances were considered. And it was found that the contributions from these nucleon resonances are non-negligible. Their contributions will significantly change the angular distributions of the $\bar{p} p \to \psi X$ reaction.

The experimental results of both the CLEO and BESIII Collaborations \cite{Ablikim:2013vtm,Alexander:2010vd} show that the nucleon resonance $N(1535)$ has a significant contribution in the decay of $\psi(3686) \rightarrow p\bar{p}\eta$. This may be because of the large coupling of $N(1535)$ to the $\eta N$ channel. As a matter of course, we will consider that $N(1535)$ may have a significant contribution in the $p\bar{p}\rightarrow \psi \eta$ reaction.
Along the above line, in this work, we will calculate the production cross sections of the process $p\bar{p}\rightarrow \psi \eta$ within the effective Lagrangian approach and also give the angular distributions, where the contributions from the nucleon pole and three $N^{*}$ states are considered. We consider the contributions from nucleon resonances $N(1520)$ $(\equiv D_{13})$ with $J^{P}=\frac{3}{2}^{-}$, as well as $N(1535)$ $(\equiv S_{11})$ and $N(1650)$ $(\equiv S_{11})$ with $J^{P}=\frac{1}{2}^{-}$, which have appreciable branching ratios for the decay into the $\eta N$ channel. On the other hand,  there are unknown model parameters, which will be determined through fitting the experimental data of $\psi\rightarrow p\bar{p} \pi^0$ and $\psi\rightarrow p\bar{p} \eta$ decays.

This article is organized as follows: First, the formalism and ingredients of $p\bar{p}\rightarrow \psi \eta$ within the effective Lagrangian approach are presented in Sec. \ref{sec2}. In Sec. \ref{sec3}, we fit the experimental data on the $\psi \to p\bar{p} \pi^0$ and $\psi \to p \bar{p} \eta$ decays to determine these unknown parameters. In Sec. \ref{sec4}, we show the numerical results and make a detailed discussion. Finally, a short summery will be given in Sec.~\ref{sec5}.

\section{Formalism and ingredients}\label{sec2}

In this section, we introduce the theoretical formalism and
ingredients for investigating the $\bar{p} p \to \psi \eta$ reaction
within the effective Lagrangian method, by including the
contributions from the nucleon pole and the low-lying nucleon
excited states that have strong couplings to the $\eta N$ channel.

\subsection{The $\bar{p} p \to \psi \eta$ reaction}

\begin{figure}[htbp]
\begin{tabular}{ccc}
\includegraphics[width=110pt]{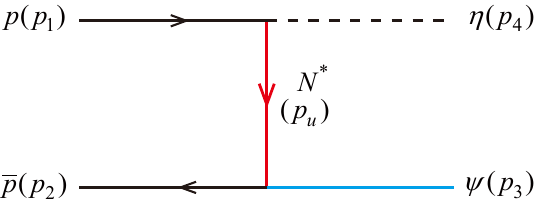}&$\quad$&
\includegraphics[width=110pt]{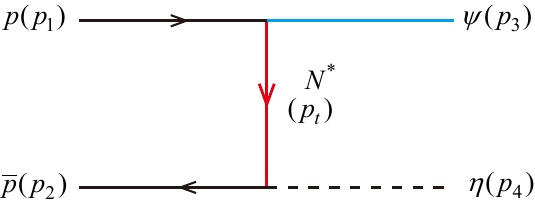}\\
(a)  &$\quad$& (b)
\end{tabular}
\caption{The Feynman diagram for the $p\bar{p}\rightarrow \psi \eta$ reaction, and (a) stands for the $u$-channel diagram; (b) stands for the $t$-channel diagram, while the $N^{*}$ represents the nucleon pole or the excited nucleon resonances.}\label{production}
\end{figure}

The production of charmonium [$\psi\equiv \psi(3686)$ and $J/\psi$]
plus a light meson $\eta$ in the low-energy $p\bar{p}$ interaction
can be achieved by exchanging intermediate nucleon and nucleon
excited states. There are two types of Feynman diagrams to depict the
$p\bar{p}\rightarrow \psi \eta$ reaction on the tree level, as shown
in Fig.~\ref{production}. It is worthy of mention that the
multipion production dominates the low-energy $\bar{p}p$
interactions, which in principle is important and its effects from
the so-called initial state interaction (ISI) should be also
considered. As discussed in
Refs.~\cite{Hanhart:1998rn,Motzke:2002fn,Baru:2002rs} for the case
of the ISI of $NN$ scattering, including such contributions, the
scattering amplitudes would be more complex due to additional model
parameters from the loop integration. Hence, we leave the
contributions from ISI of $\bar{p}p$ to further studies when more
precise experimental data become available.

In this calculation, we use the effective interaction Lagrangian
densities for each vertexs in Fig.~\ref{production}. For the $\psi
N\bar{N}$ and $\eta N\bar{N}$ vertices, we use the effective
Lagrangians as
\begin{eqnarray}\label{npi}
\mathcal {L}_{\eta NN}&=&-ig_{\eta NN}\bar{N}\gamma_5 \eta N ,  \\
\mathcal {L}_{\psi NN}&=&-g_{\psi NN}\bar{N}\gamma_\mu V^\mu N,
\end{eqnarray}
where $V^\mu$ donates the vector field of $\psi$.

For the $N^* N \eta$ and $\psi N^* \bar N$ vertices, we adopt the Lagrangian densities as used in
Refs.~\cite{Tsushima:1996xc,Tsushima:1998jz,Ouyang:2009kv,Wu:2009md,Cao:2010km,Cao:2010ji,Zou:2002yy}:
\begin{eqnarray}
\mathcal {L}_{\eta NR_{S_{11}}}&=&-g_{\eta NR_{S_{11}}}\bar{N} \eta R_{S_{11}}+H.c.,  \\
\mathcal {L}_{\eta NR_{D_{13}}}&=&-\frac{g_{\eta
NR_{D_{13}}}}{m_N^2}\bar{N}\gamma_5\gamma^\mu\partial_\mu\partial_\nu \eta R^\nu_{D_{13}}+H.c., \\
\mathcal {L}_{\psi NR_{S_{11}}}&=&-g_{\psi NR_{S_{11}}}\bar{N}\gamma_5 \gamma_\mu V^\mu R_{S_{11}}+H.c.,  \\
\mathcal {L}_{\psi NR_{D_{13}}}&=&-g_{\psi NR_{D_{13}}}\bar{N}V_\mu
R^\mu_{D_{13}}+H.c. ,\label{nsv}
\end{eqnarray}
where $R$ denotes the $N^*$ field.

Then, we can write the scattering amplitudes of the process $p\bar{p}\rightarrow \psi \eta $ as,

\begin{widetext}
\begin{eqnarray}
\label{1/2+}
\mathcal{M}_{N}&=&ig_{\eta N R_{P_{11}}}g_{\psi N R_{P_{11}}}\bar{v}(p_2)\varepsilon^{\mu}(p_3)
\left[\gamma_{\mu}G^{\frac{1}{2}}(p_u)\gamma_{5}\mathcal{F}(u)
+\gamma_{5}G^{\frac{1}{2}}(p_{t})\gamma_{\mu}\mathcal{F}(t)\right]u(p_{1}),\\
 \label{1/2-}
\mathcal{M}_{S_{11}}&=&g_{\eta N R_{S_{11}}}g_{\psi N R_{S_{11}}}\bar{v}(p_2)\varepsilon^{\mu}(p_3)
\left[\gamma_{5}\gamma_{\mu}G^{\frac{1}{2}}(p_{u})\mathcal{F}(u)
+G^{\frac{1}{2}}(p_{t})\gamma_{5}\gamma_{\mu}\mathcal{F}(t)\right]u(p_{1}),\\
\label{3/2-}
\mathcal{M}_{D_{13}}&=&\frac{g_{\eta N R_{D_{13}}}}{m_{N}^{2}}g_{\psi N D_{13}}\bar{v}(p_2)\varepsilon^{\mu}(p_3)
\left[G_{\mu\nu}^{\frac{3}{2}}(p_{u})(i\gamma_{5}\slashed{p}_{4})(ip_{4}^{\nu})\mathcal{F}(u)
+(i\gamma_{5}\slashed{p}_{4})(ip_{4}^{\nu})G_{\mu\nu}^{\frac{3}{2}}(p_{t})\mathcal{F}(t)\right]u(p_{1}),
\end{eqnarray}
\end{widetext}
where $u=p_{u}^{2}=(p_1-p_4)^{2}=(p_3-p_2)^{2}$, $t=p_{t}^{2}=(p_1-p_3)^{2}=(p_4-p_2)^{2}$. The $\mathcal{F}(u)$ and $\mathcal{F}(t)$ stand for the form factors of the $u$ and $t$ channels, respectively. Besides this, we adopt the expression as used in Refs.~\cite{Feuster:1997pq,Haberzettl:1998eq,Yoshimoto:1999dr,Oh:2000zi}:
\begin{eqnarray}
\mathcal{F}(u/t)=\frac{\Lambda_{N^{*}}^{4}}{\Lambda_{N^{*}}^{4}+(u/t-m_{N^{*}}^{2})},
\end{eqnarray}
where the cutoff parameter $\Lambda_{N^{*}}$ can be parametrized as
\begin{eqnarray}
\Lambda_{N^{*}}=m_{N^{*}}+\beta \Lambda_{QCD},
\end{eqnarray}
with $\Lambda_{QCD} = 220$ MeV, and the $\beta$ will be determined by fitting the experimental data on the $\psi(3686) \to p \bar{p} \eta$ decay.

The Breit-Wigner form of the propagator $G^{J}(p)$ for the $J=\frac{1}{2}$ and $J=\frac{3}{2}$ can be written as \cite{Huang:2005js}
\begin{eqnarray}
G^{\frac{1}{2}}(p)&=&i\frac{\slashed{p}+m_{N^{*}}}{p^{2}-m_{N^{*}}^{2}+im_{N^{*}\Gamma_{N^{*}}}},\\ \nonumber
G^{\frac{3}{2}}_{\mu\nu}(p)&=&i\frac{\slashed{p}+m_{N^{*}}}{p^{2}-m_{N^{*}}^{2}+im_{N^{*}\Gamma_{N^{*}}}}\Bigg[-g_{\mu\nu}+\frac{1}{3}\gamma_{\mu}\gamma_{\nu}\\ \nonumber
&&+\frac{1}{3m_{N^{*}}}(\gamma_{\mu}p_{\nu}-\gamma_{\nu}p_{\mu})+\frac{2p_{\mu}p_{\nu}}{3m_{N^{*}}^{2}}\Bigg].
\end{eqnarray}

Note that we take the energy-dependent form for the decay width $\Gamma_{N^*}$ of $N(1535)$ resonance, and we take the energy-dependent form, which is given by~\cite{Xie:2017erh}
\begin{eqnarray}
\Gamma_{N^*} (q^2) &=& \Gamma_{N^* \to \pi N}(q^2) + \Gamma_{N^* \to
\eta N}(q^2) + \Gamma_0,   \label{Eq:GamrNstarq2} \\
\Gamma_{N^* \to \pi N}(q^2) \! &=& \!\! \frac{3g^2_{N^*N\pi}}{4\pi} \frac{\sqrt{|\vec{p}_{N\pi}|^2 + m^2_p} + m_p}{\sqrt{q^2}} |\vec{p}_{N\pi}|, \\
\Gamma_{N^* \to \eta N}(q^2) \! &=& \!\! \frac{g^2_{N^*N\eta}}{4\pi}
\frac{\sqrt{|\vec{p}_{N\eta}|^2 + m^2_p} + m_p}{\sqrt{q^2}}
|\vec{p}_{N\eta}|,
\end{eqnarray}
with
\begin{eqnarray}
|\vec{p}_{N\pi}| &=&
\frac{\lambda^{1/2}(q^2,m^2_p,m^2_{\pi})}{2\sqrt{q^2}}, \\
|\vec{p}_{N\eta}| &=&
\frac{\lambda^{1/2}(q^2,m^2_p,m^2_{\eta})}{2\sqrt{q^2}},
\end{eqnarray}
where $\lambda$ is the K\"allen function with $\lambda(x,y,z) =
(x-y-z)^2 -4yz$. We take $g_{\pi N^*N}=0.62$ and
$g_{\eta N^*N} = 1.85$, which are determined from the partial widths of $N(1535)$ decay to $N\pi$ and $N\eta$. With
these values we can get $\Gamma_{N^* \to N\pi} = 54.9$ MeV and
$\Gamma_{N^* \to N\eta} = 55.1$ MeV if we take $\sqrt{q^2} = 1524$
MeV. To  agree with experimental results, we choose $\Gamma_0 = 19$ MeV for
$\Gamma_{N^*}(q^2 ) = 130$ MeV.
Here, the mass and width of $N(1535)$ are adopted in Ref.\cite{Ablikim:2013vtm}.

The other coupling constants in the above Lagrangian densities can be also determined from their partial decay widths. The obtained numerical results for these relevant coupling constants are listed in Tables \ref{parameters2S} and \ref{parameters1S}. The coupling constants $g_{\psi NN^*}$ are obtained from the decay process of $\psi \to \bar{N}N^* + N \bar{N}^* \to p\bar{p}\pi^0$, while the coupling constant $g_{J/\psi \bar{p}p} = 1.63$ is extracted from $J/\psi \to \bar{p}p$. In addition, we will discuss the coupling constants $g_{\eta NN}$ and $g_{\psi(3686) NN}$ below.

\begin{table}[htbp]
\caption{The coupling parameters $ g^{J/\psi\pi}_{N^{*}}$ are estimated from the branching fraction (B.F.) of each intermediate nucleon resonance of $J/\psi\rightarrow N \bar{N}^{*}+ N^{*}\bar{N}\rightarrow \bar{p}p\pi^{0}$ (second column), the width of $J/\psi$ is 92.9 KeV. The last column is the parameter of $g^{J/\psi\eta}_{N^{*}}$ which are estimated by the formula $g^{J/\psi\eta}_{N^{*}}=g_{\eta N^{*}N}\frac{g^{J/\psi\pi}_{N^{*}}}{g_{\pi N^{*}N}}$.}\label{parameters1S}
\begin{tabular}{cccc}
\toprule[1pt]
\midrule[1pt]
$N^{*}$s & $\rm{B.F.}_{(J/\psi\rightarrow \bar{p}p\pi^{0})}$($\times 10^{-5}$)& $g^{J/\psi\pi}_{N^{*}}$ ($\times 10^{-3}$) & $g^{J/\psi\eta}_{N^{*}}$ ($\times 10^{-3}$) \\
\midrule[1pt]
$N(940)$&-&21.87 & 14.56\\
$N(1520)$&7.96&4.36 & 4.87 \\
$N(1535)$&7.58&0.85 & 2.55\\
$N(1650)$&9.06&0.99 & 1.42\\
\midrule[1pt]
\bottomrule[1pt]
\end{tabular}
\end{table}

\begin{table*}[htbp]
\centering
\caption{The coupling parameters of $g_{\pi N^{*}N}$ (seventh column) and $g_{\eta N^{*}N}$ (eighth column) are estimated from the branching ratios of $N^{*}\rightarrow N\pi$ and $N^{*}\rightarrow N\eta$ respectively.
The ninth column gives the parameter $g_{N^{*}}^{\psi^{'} \pi}$, which is estimated from the branching ratio of $\psi(3686)\rightarrow N \bar{N}^{*}+ N^{*}\bar{N}\rightarrow \bar{p}p\eta$; the last column gives $g_{N^{*}}^{\psi^{'} \eta}$, which is calculated from the relationship that exists in $g_{\pi N^{*}N}$, $g_{\eta N^{*}N}$, $g_{N^{*}}^{\psi^{'} \pi}$ and $g_{N^{*}}^{\psi^{'}\eta}$. In this table, relevant experimental data are adopted from PDG~\cite{Tanabashi:2018oca}, but for the mass and total width of $N(1535)$ we adopt the values given in Ref.~\cite{Ablikim:2013vtm}.}\label{parameters2S}
\begin{tabular}{cccccccccc}
\toprule[1pt]
\midrule[1pt]
$N^*$ & Mass (GeV) & $\Gamma$ (GeV) & $\rm{B.F.}_{(N^{*}\rightarrow N\pi)}$ & $\rm{B.F.}_{(N^{*}\rightarrow N\eta)}$& $\rm{B.F.}_{(\psi(3686)\rightarrow p\bar{p}\pi^{0})}$($\times 10^{-5}$)& $g_{\pi N^{*}N}$&$g_{\eta N^{*}N}$& $g^{\psi(3686)\pi}_{N^{*}}$ ($\times 10^{-3}$) & $g^{\psi(3686)\eta}_{N^{*}}$ ($\times 10^{-3}$) \\
\midrule[1pt]
$N(1520)$& 1.515&0.110&0.6&$8.0\times10^{-4}$&0.64&4.09&4.55 & 1.09 & 1.22 \\
$N(1535)$&1.524&0.130&0.42&0.425&2.47&0.62&1.85&0.51 & 1.53\\
$N(1650)$&1.650&0.125&0.6&0.25&3.76&0.68&0.98&0.62 & 0.89\\
\midrule[1pt]
\bottomrule[1pt]
\end{tabular}
\end{table*}

Finally, in the center-of-mass frame denoted by the $cm$ superscript, the differential cross section of $p\bar{p}\rightarrow \psi \eta$ process can be written as
\begin{eqnarray}
\frac{d\sigma}{dcos\theta}=\frac{1}{32\pi s}\frac{\left|\overrightarrow{p}_{3}^{cm}\right|}{\left|\overrightarrow{p}_{1}^{cm}\right|} \overline{\left|\mathcal{M}_{tot}\right|^{2}}, \label{PCformula}
\end{eqnarray}
where $\theta$ is the scattering angle of outgoing $\eta$ relative to the  direction of the antiproton beam in the center-of-mass frame, while $\overrightarrow{p}_{1}^{cm}$ and $\overrightarrow{p}_{3}^{cm}$ are the three-momenta of the proton and $\psi$ in the center-of-mass frame, respectively. $\mathcal{M}_{tot}$ is the total invariant scattering amplitude of the $\bar{p} p \to \psi \eta$ reaction, which can be written as
\begin{eqnarray}
\label{Mtot1}
\mathcal{M}_{tot}^{\bar{p}p\rightarrow \psi\eta}=\mathcal{M}_{N}+\sum\limits_{N^{*}} \mathcal{M}_{N^{*}}e^{-i\phi_{N^{*}}},
\end{eqnarray}
where $\mathcal{M}_{N}$ and $\mathcal{M}_{N^{*}}$ are the contributions from the nucleon pole and the nucleon resonances, respectively. Besides this, we introduce the relative phase $\phi_{N^{*}}$ between $\mathcal{M}_{N^{*}}$ and $\mathcal{M}_{N}$.

\subsection{Determine the model parameters from the analysis of the $\psi(3686) \rightarrow \bar{p}p \eta$ decay} \label{sec3}

\begin{figure}[htbp]
\begin{tabular}{ccc}
\includegraphics[width=110pt]{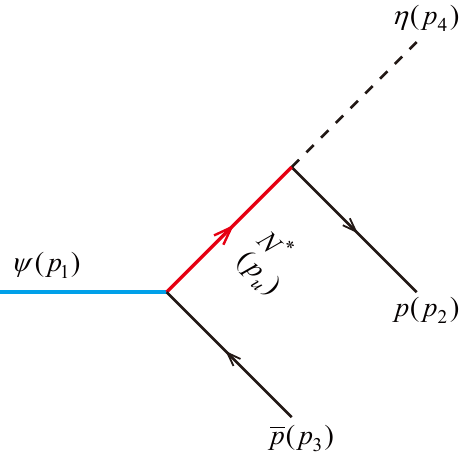}&$\quad$&
\includegraphics[width=110pt]{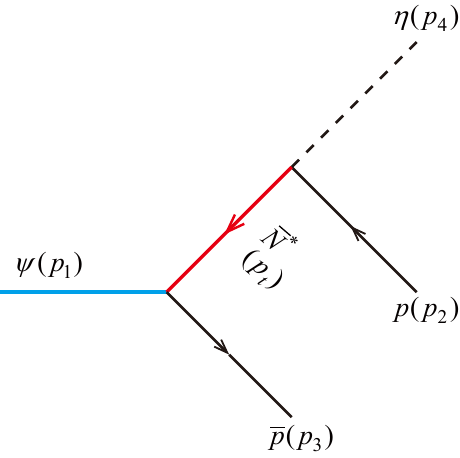}\\
(a)  &$\quad$& (b) \\
\end{tabular}
\caption{The Feynman diagram of $\psi \rightarrow \bar{p}p \eta$ through the nucleon pole and the nucleon excited states. The panel (a) shows $u$-channel exchange; panel (b) is $t$-channel exchange.}\label{decay}
\end{figure}

On the tree level, the process $\psi \rightarrow  \bar{p}p \eta$ is described by the Feynman diagrams as shown in Fig.~\ref{decay}.

With the effective interaction Lagrangian densities given above, we can easily obtain the decay width of $\psi \rightarrow \bar{p}p \eta$, which can be written as
\begin{eqnarray}\label{dfferentialwidth}
d\Gamma=\frac{1}{(2\pi)^{5}}\frac{1}{16M^{2}}\overline{|M_{tot}^{\psi\rightarrow \bar{p}p\eta}|^{2}}|\vec{p}_{2}^{*}||\vec{p}_{3}|d\Omega_{2}^{*}d\Omega_{3}dm_{p\eta},
\end{eqnarray}
where $\vec{p}_{2}^{*}$ ($\Omega_{2}^{*}$) stands for the three-momentum (solid angle) of the proton in the rest frame of the $p$ and $\eta$ system, $\vec{p}_{3}$ ($\Omega_{3}$) is the three-momentum (solid angle) of the antiproton in the rest frame of $\psi$, and $m_{p\eta}$ is the invariant mass of the $p$ and $\eta$ system. On the other hand, the amplitude $\mathcal{M}_{tot}^{\psi\rightarrow \bar{p}p\eta}$ is easily obtained just by applying the substitution to $\mathcal{M}_{tot}^{p\bar{p}\rightarrow \psi\eta}$:
$p_{1}\rightarrow -p_{3}$, $p_{2}\rightarrow -p_{2}$, $p_{3}\rightarrow -p_{1}$, $p_{t}\rightarrow -p_{u}$, $\varepsilon(p_{3})\rightarrow\varepsilon(-p_{1})$, $u(p_{1})\rightarrow v(-p_{3})$, $\bar{v}(p_{2})\rightarrow \bar{u}(-p_{2})$.

The $\psi(3686)\rightarrow  \bar{p}p \eta$ decay is experimentally studied by the CLEO and BESIII Collaborations~\cite{Alexander:2010vd,Ablikim:2013vtm}, and they found that most contributions are from nucleon excited state $N(1535)$, which has large coupling to the $N\eta$ channel. However, since $N(1520)$ and $N(1650)$ have significant couplings to the $N\eta$ channel, in this work, we will also take their contributions into account. Then, we perform five-parameter ($g^{\psi(3686)\eta}_N \equiv g_{\psi(3686) NN} \times g_{\eta NN}$, $\phi_{N(1520)}$, $\phi_{N(1535)}$, $\phi_{N(1650)}$ and $\beta$)~\footnote{Note that we take the same parameter $\beta$ for all the nucleon resonances that we considered, and all the other parameters are shown in Table~\ref{parameters2S}.} $\chi^2$ fits to the experimental data~\cite{Ablikim:2013vtm} on the $p\eta$ invariant mass distributions for the $\psi(3686)\rightarrow N\bar{N}^* + N^*\bar{N} \to  \bar{p}p \eta$ decay.

\begin{figure}[htbp]
\includegraphics[width=240pt]{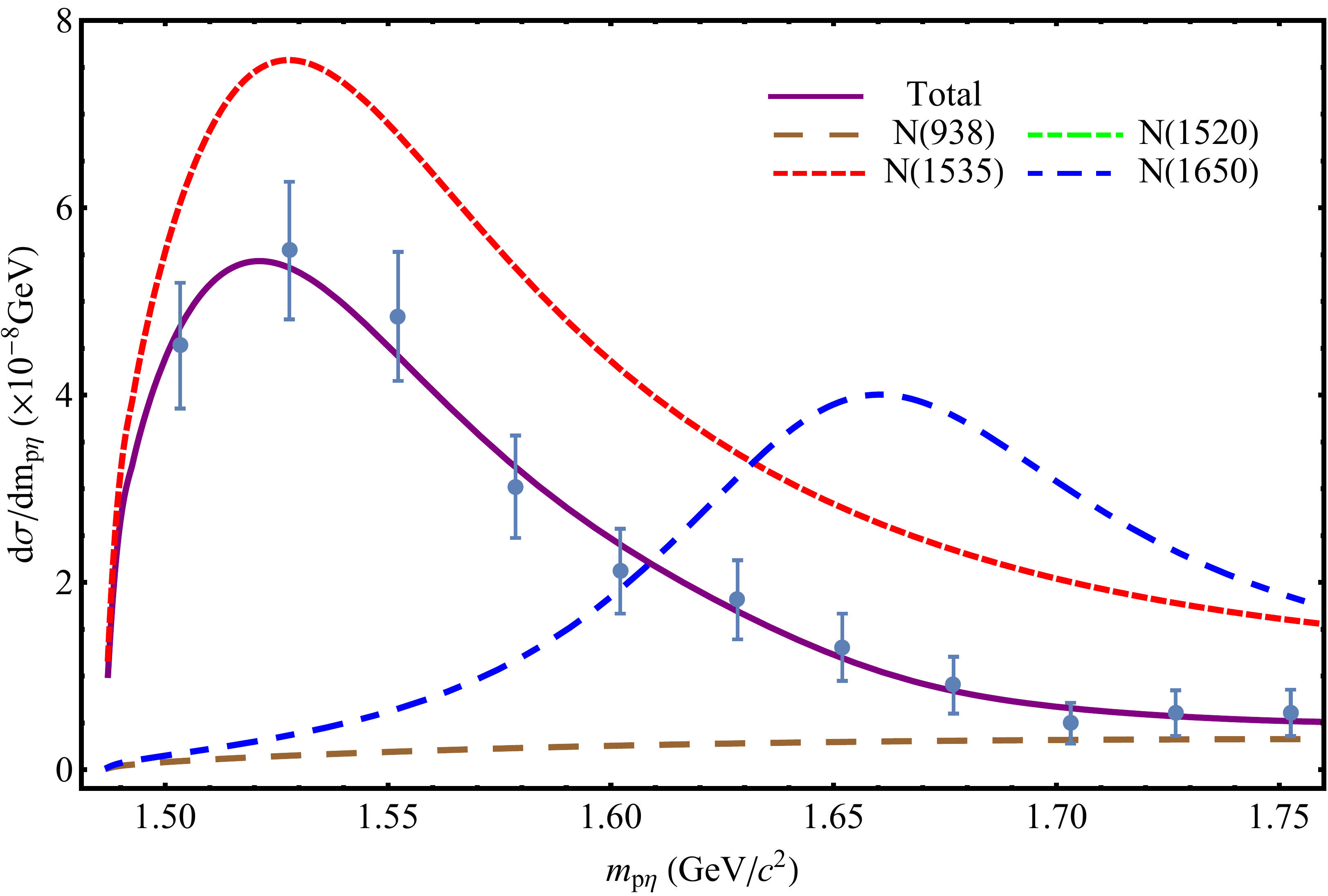}
\caption{The fitted $p\eta$ invariant mass distributions of the
process $\psi(3686)\rightarrow N \bar{N}^{*}+ N^{*}\bar{N}\rightarrow \bar{p}p\eta$ with experiment data taken from Ref.~\cite{Ablikim:2013vtm}. The solid purple line stands for the total contributions, and the other dashed lines show the contributions from the nucleon pole and different
 nucleon resonances.}\label{fittingresulttot}
\end{figure}

The fitted parameters are: $\phi_{N(1520)} = 0.14 \pm 0.08$, $\phi_{N(1535)} = 1.76 \pm 0.06$, $\phi_{N(1650)} = 4.63 \pm 0.05$, $\beta = 3.70 \pm 0.82$, and $g^{\psi(3686) \eta}_N = (8.45 \pm 1.10) \times 10^{-3}$. The resultant $\chi^{2}/dof$ is $0.39$. The best-fitted results are shown in Fig.~\ref{fittingresulttot}, compared with the experimental data. One can see that we can describe the experimental data quite well. $N(1535)$ gives the dominant contribution below $m_{\eta p} = 1.6$ GeV, and the contribution from $N(1650)$ is also significant, while the other contributions are quite small. Furthermore, there are strong interferences between $N(1535)$ and $N(1650)$, which make the peak of $N(1650)$ disappear in the total results.

Besides, if we take $g_{\psi(3686) NN} =9.4\times 10^{-4} $ and $g^{\psi(3686) \eta}_N = (8.45 \pm 1.10) \times 10^{-3} $, which was obtained from the $\chi^2$ fits, we can easily obtain $g_{\eta NN}= 8.99 \pm 1.17$, which is in the range of many other theoretical results on it~\cite{Kirchbach:1996kw,Zhu:2000eh,Faldt:2001uz,Cottingham:1973wt,Lacombe:1980dr,Nagels:1978sc,Machleidt:2000ge,Downum:2006re}.

\section{The total cross sections and angular distributions of $p\bar{p}\rightarrow \psi \eta$}\label{sec4}

In this section, we show theoretical results on the total cross sections and angle distributions of the $p\bar{p}\rightarrow \psi \eta$ reaction near reaction threshold.

\subsection{The total cross sections and angular distributions of $p\bar{p}\rightarrow \psi(3686) \eta$}\label{Sec4A}

\begin{figure}[htbp]
\includegraphics[width=245pt]{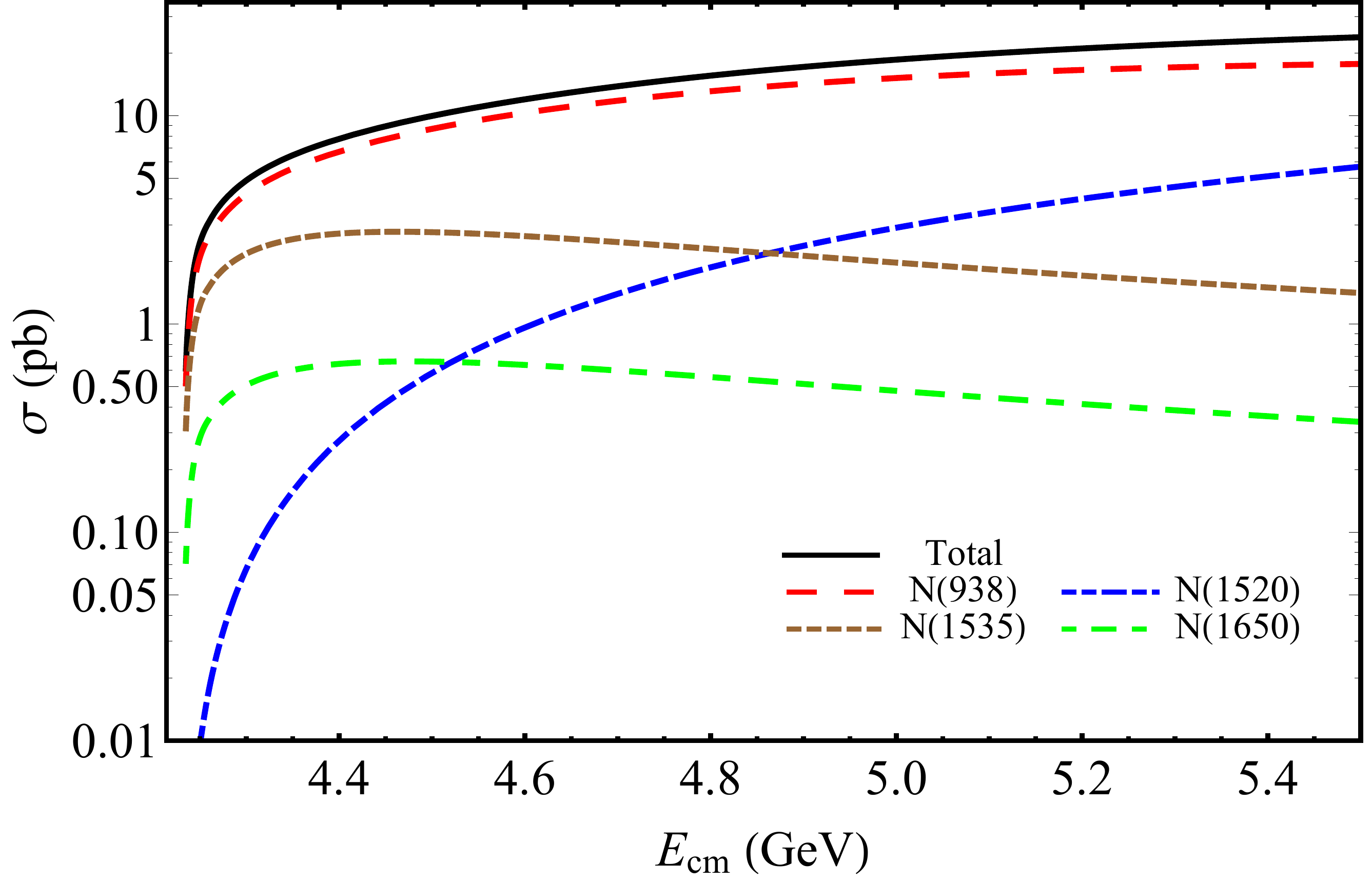}
\caption{The total cross sections of the $p \bar{p}\rightarrow\psi(3686) \eta$ reaction. The black line is the total contributions, while the other lines stand for the contributions from different $N^{*}$ resonances.}\label{production2S}
\end{figure}

In Fig. \ref{production2S}, we show the numerical cross sections of $p\bar{p}\rightarrow \psi(3686)\eta$ as a function of the center-of-mass energy $E_{cm} = \sqrt{s}$. It is shown that the nucleon pole contribution is predominant in the whole energy region, but the contributions of the $N^{*}$ resonances gradually become significant when the $E_{cm}$ is increasing, especially the contribution from $N(1520)$. The contributions from $N(1535)$ and $N(1650)$ are mainly reflected in the forepart and begin to decrease around $E_{cm}=4.4$ GeV.
Although the contribution of $N(1535)$ is very predominant in the decay process $\psi(3686)\rightarrow \bar{p}p\eta$,  its contribution to $p\bar{p}\rightarrow \psi(3686) \eta$ is not so important. The contributions of the $N^*$ resonances are suppressed due to the highly off-shell effect of their propagators in the $t$ and $u$ channels.

\begin{figure}[htbp]
\begin{tabular}{cccc}
\includegraphics[width=115pt]{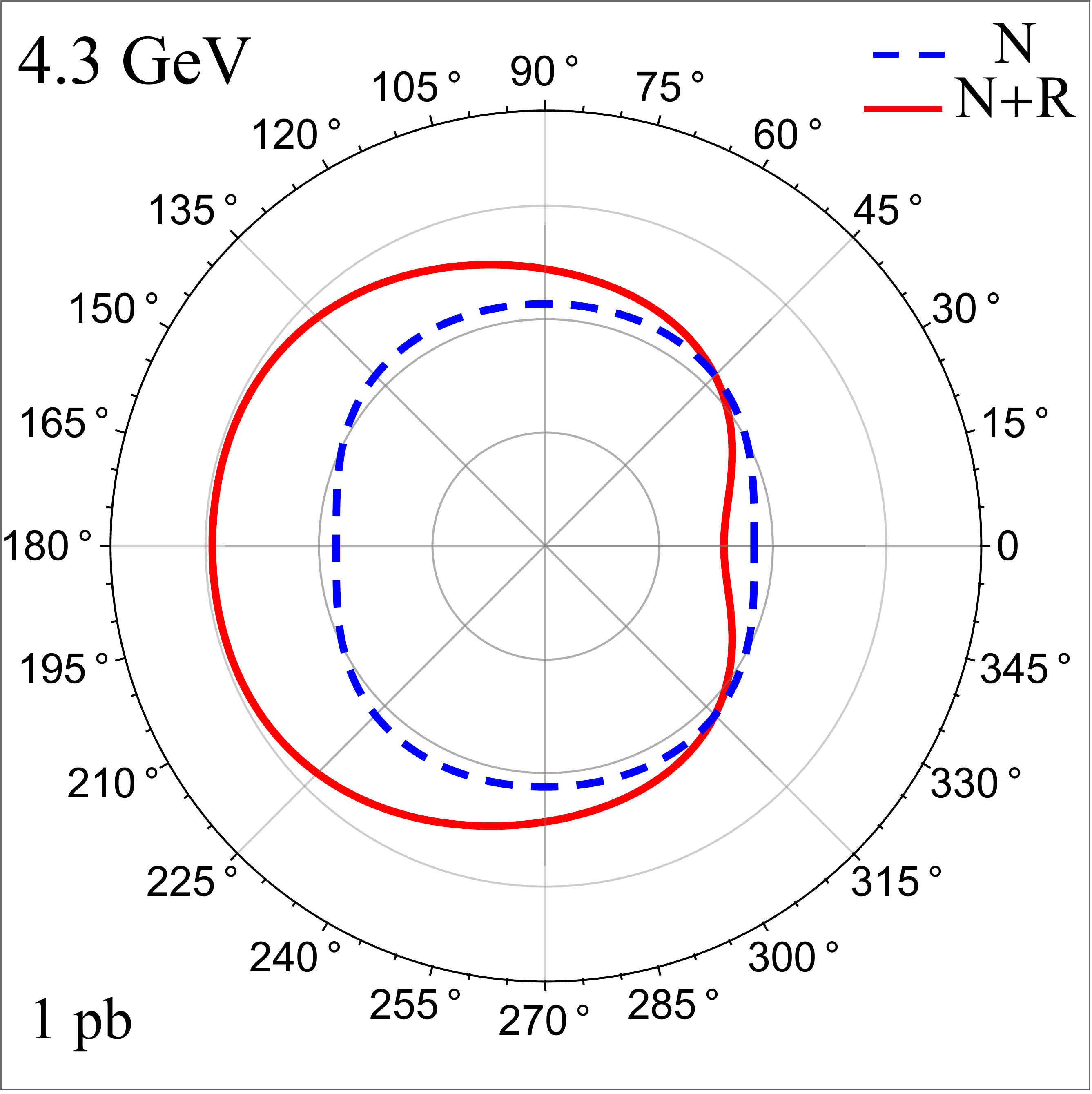}&\includegraphics[width=115pt]{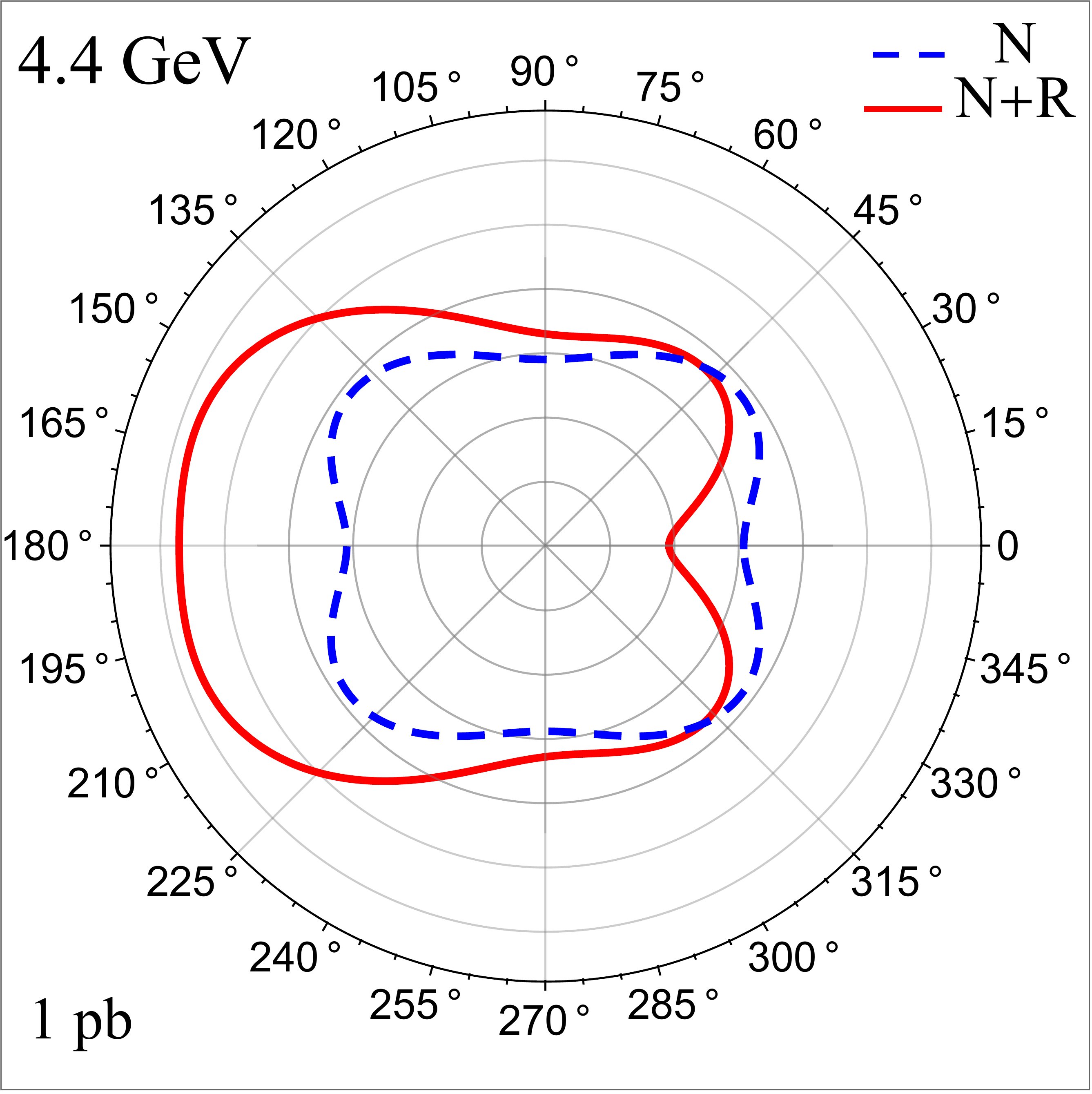}\\
\includegraphics[width=115pt]{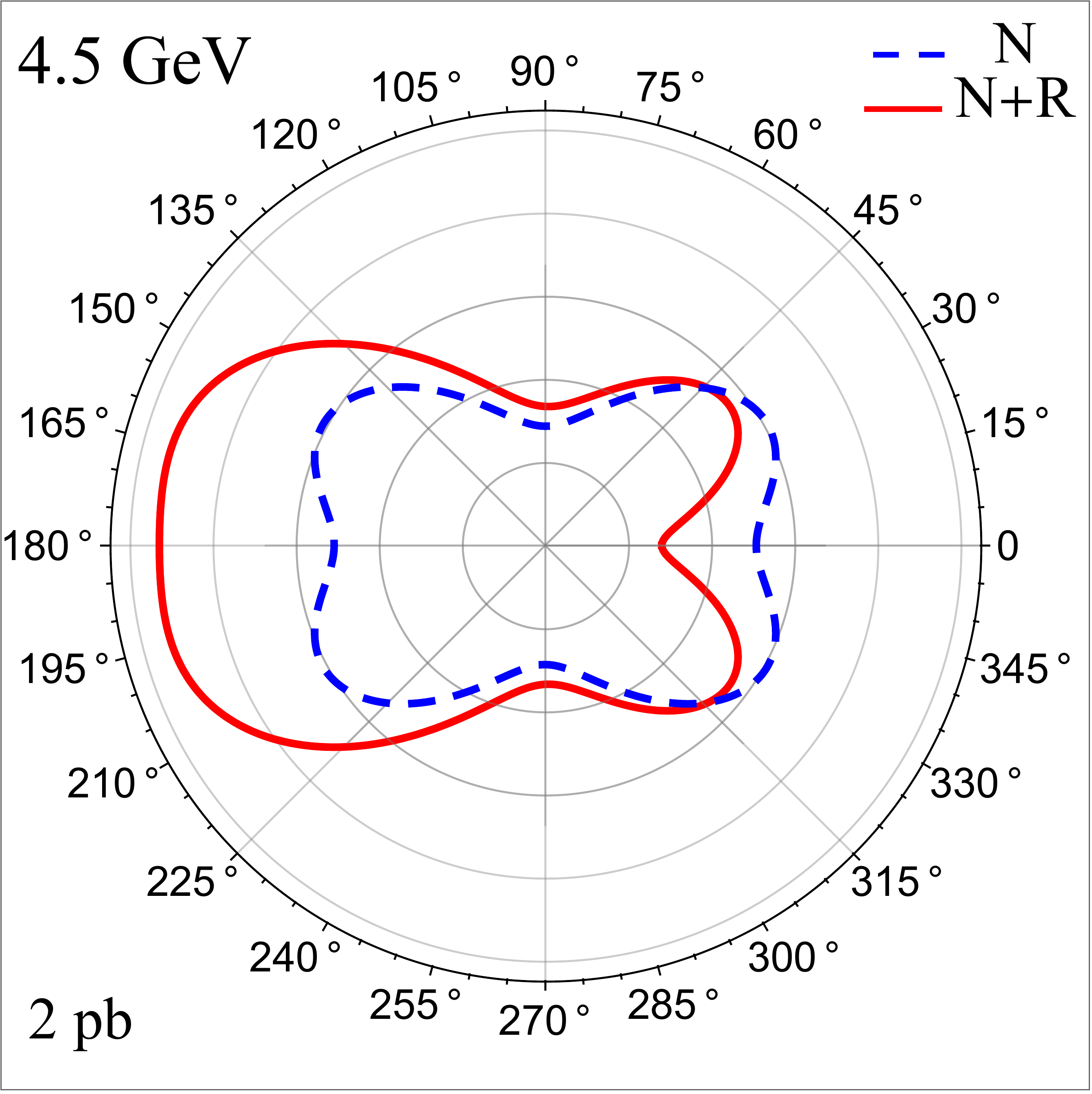}&\includegraphics[width=115pt]{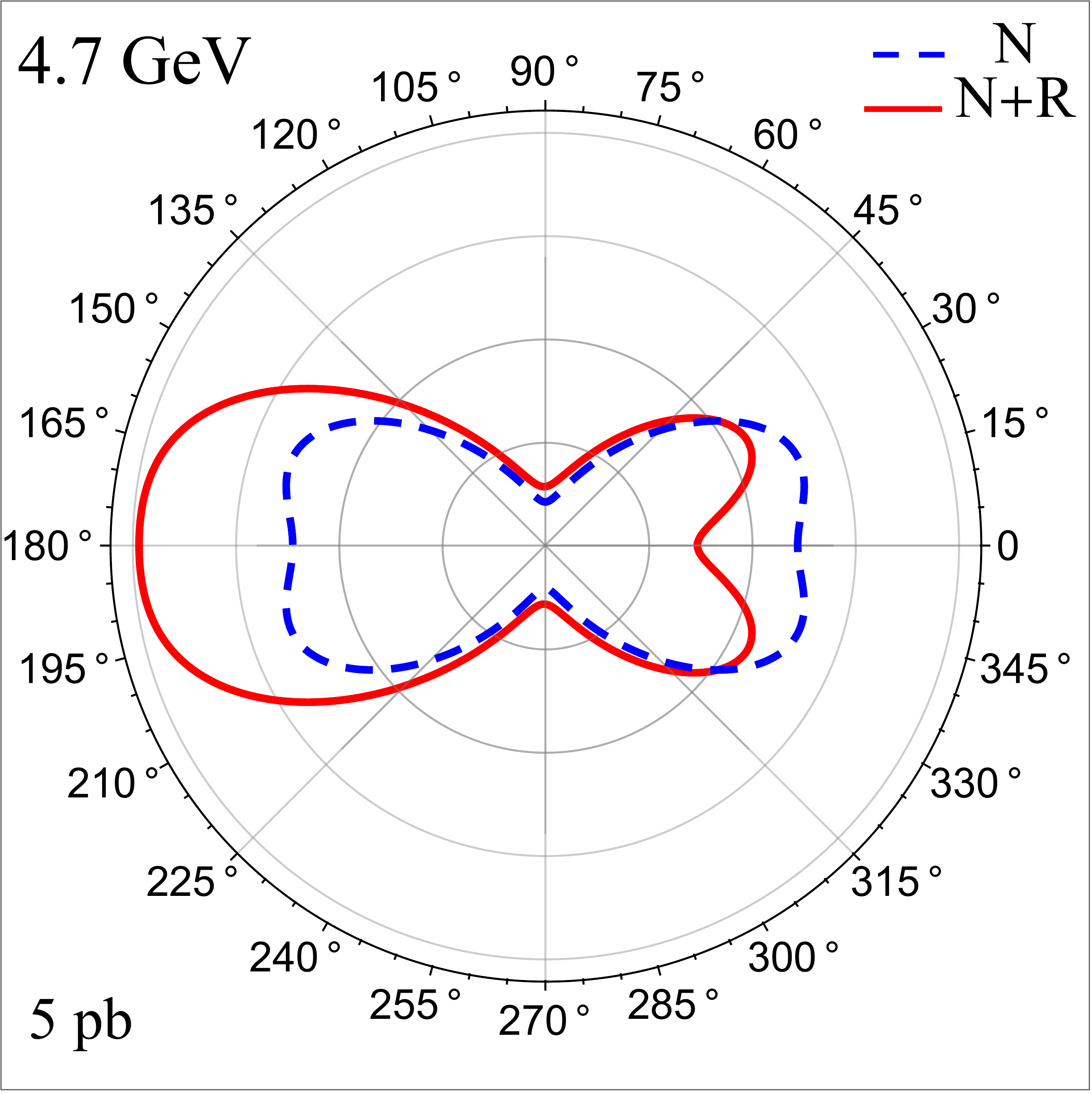}\\
\includegraphics[width=115pt]{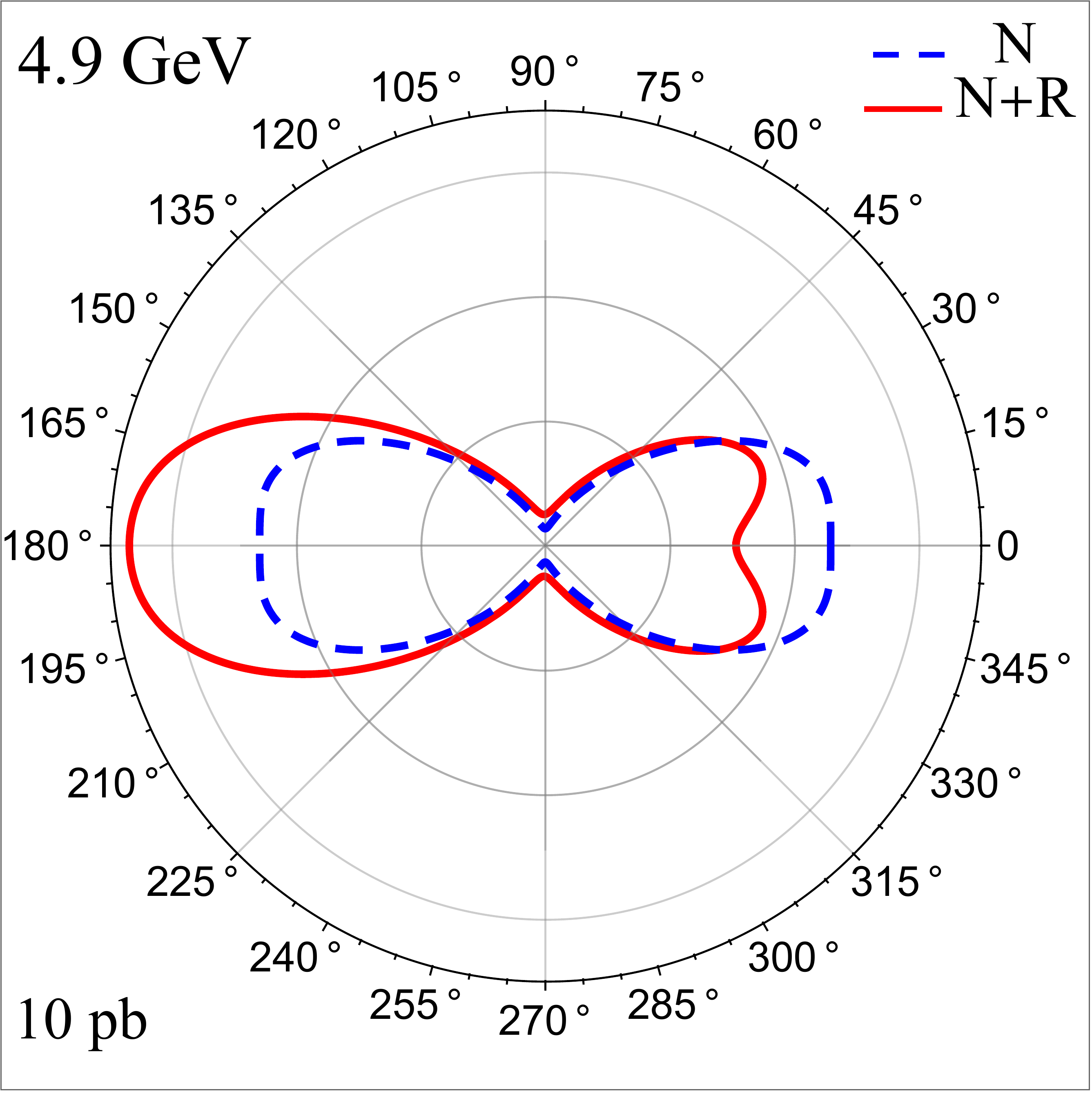}&\includegraphics[width=115pt]{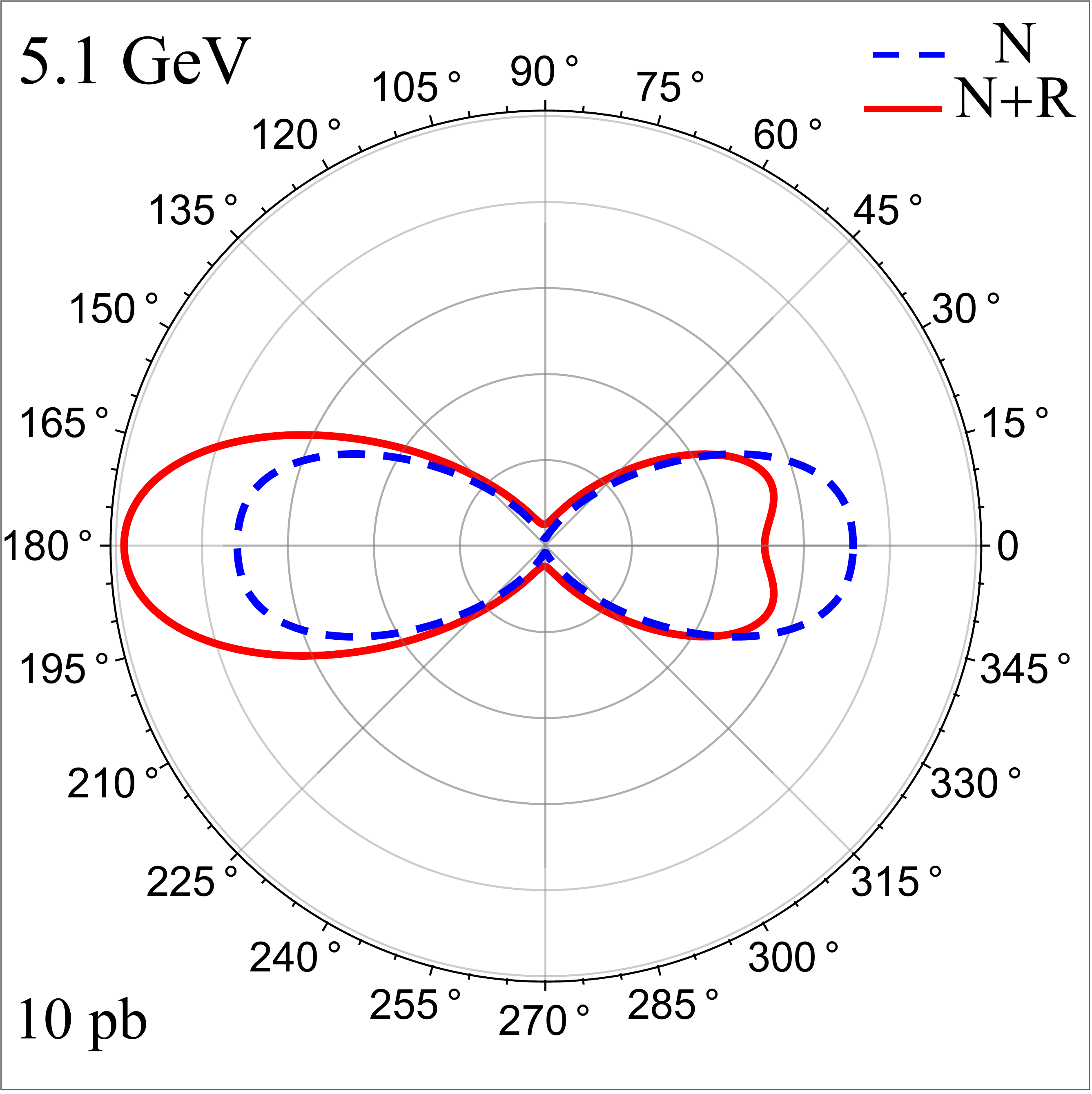}\\
\includegraphics[width=115pt]{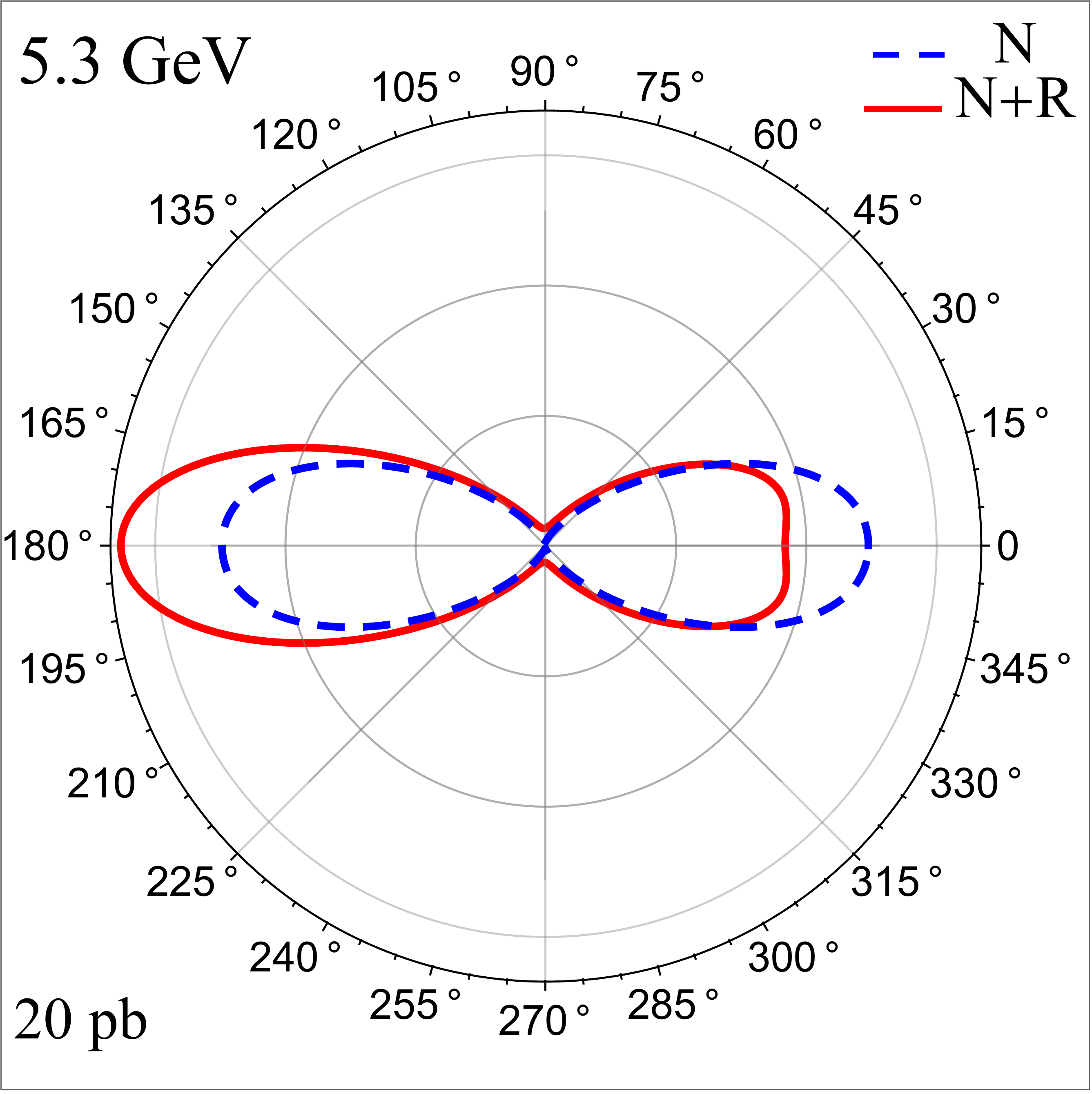}&\includegraphics[width=115pt]{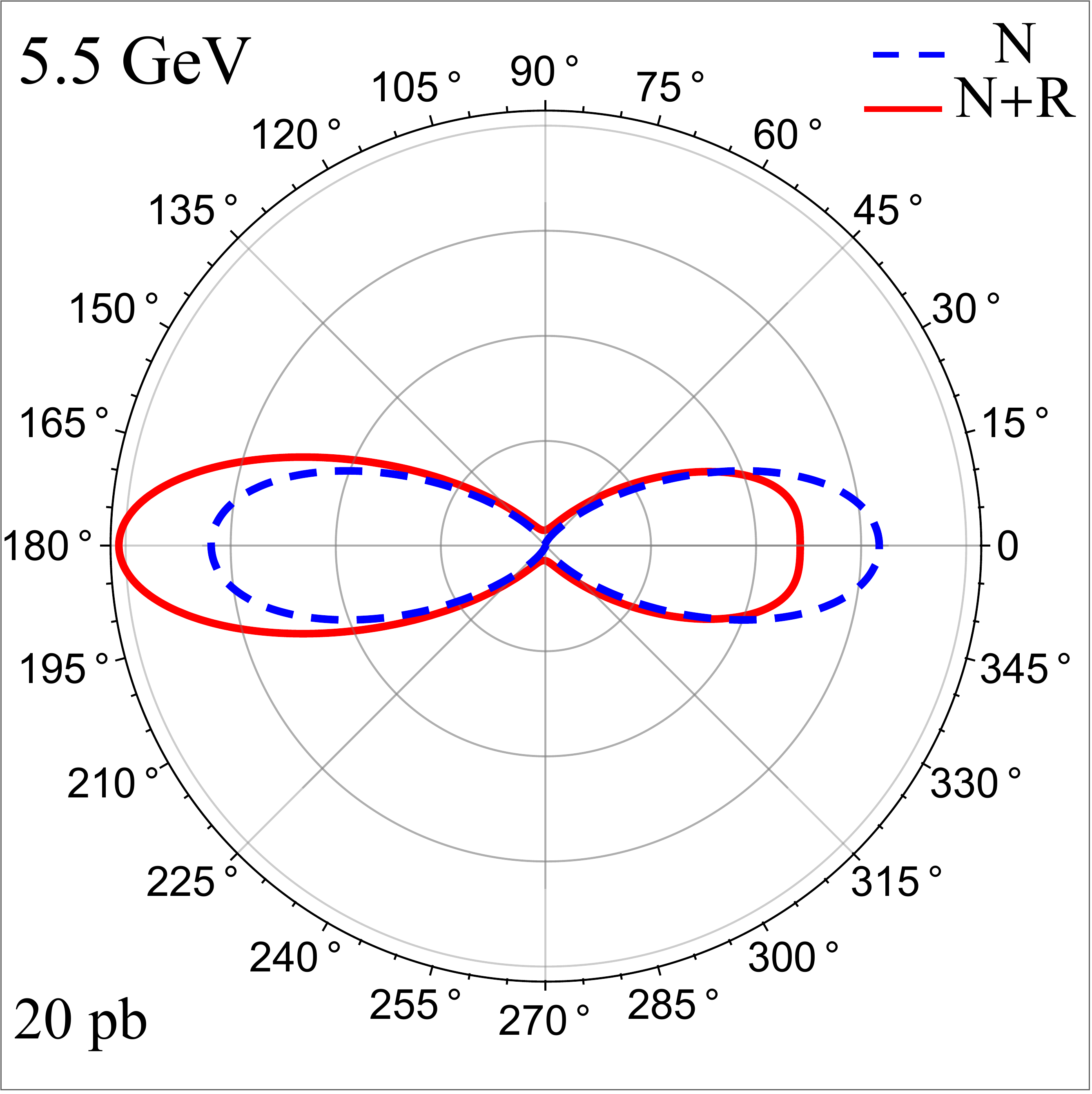}
\end{tabular}
\caption{The angular distribution of the process $p \bar{p}\rightarrow \psi^{\prime}\eta$ at different values of $E_{cm}$ (labeled in the top left corner of each panel). The red solid line is the result of considering the total contribution from the nucleon pole and nucleon resonances, while the blue dashed line is the result of only considering the contribution of the nucleon pole. Each gray concentric circles denotes a specific same value of $d \sigma/d \cos \theta$; from inside to outside, two neighboring concentric circles have the same value added (labeled in the bottom left corner of each panel).}\label{angulardis2S}
\end{figure}

We also calculate the angular distribution of the $p\bar{p}\rightarrow \psi(3686)\eta$ reaction at $E_{cm}=4.3$, 4.4, 4.5, 4.7, 4.9, 5.1, 5.3 and 5.5 GeV. The numerical results of $d \sigma/d \cos \theta$ as a function of $\theta$ are shown in Fig. \ref{angulardis2S}, where the red solid line stands for the total contribution, and the blue dashed line is the result of only considering the contribution of the nucleon pole.
Each gray concentric circle denotes a specific value of $d \sigma/d \cos \theta$, and these concentric circles are evenly spaced from inside to outside, with the difference value is labeled in the bottom left corner of each panel. From Fig. \ref{angulardis2S}, one sees that the blue dashed lines are symmetric with respect to $\theta=90\degree$ or $\theta=270\degree$, which is because the contributions from $u$ channel and $t$ channel have the same weights for all $E_{cm}$ if only the contribution from the nucleon pole is considered. However, the shapes of the red solid lines are not symmetric with respect to $\theta=90\degree$. The symmetry behavior of the angular distribution is helpful to identify the role of excited nucleon resonances in the $p\bar{p}\rightarrow \psi(3686)\eta$ reaction in future $\bar{\mathrm{P}}$ANDA experiments.

\subsection{The total cross sections and angular distributions of $p\bar{p}\rightarrow J/\psi \eta$}\label{Sec4A}

\subsubsection{The $J/\psi \rightarrow  p\bar{p} \eta$ decay}

\begin{figure}[htbp]
\includegraphics[width=240pt]{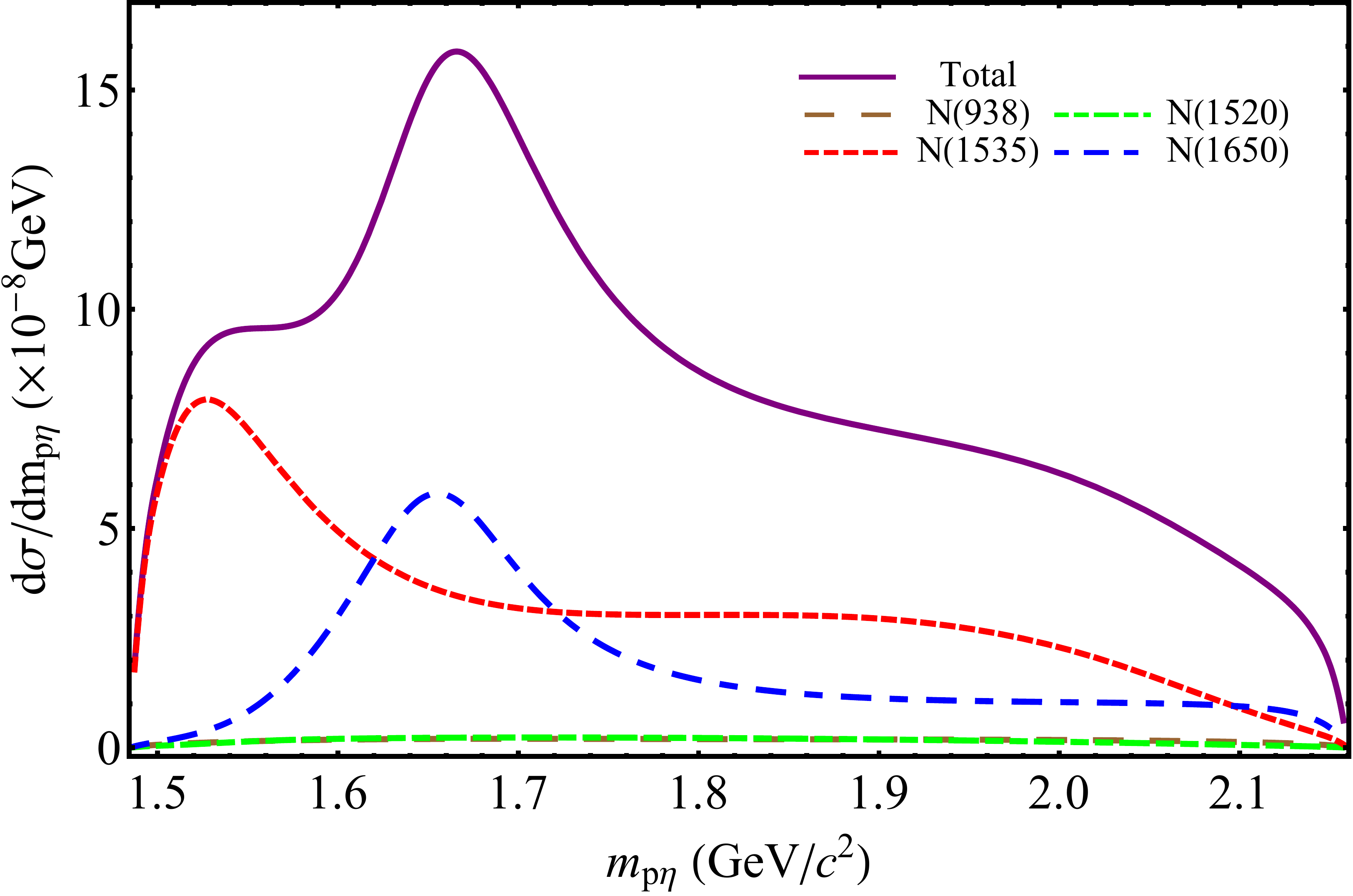}
\caption{The $p\eta$ invariant mass spectrum of the
process $J/\psi\rightarrow p\bar{p}\eta$. The solid purple line stands for the total
contributions, and the other dashed lines show the contributions from the nucleon pole and different
 nucleon resonances.}\label{fittingresulttot1S}
\end{figure}

\begin{figure}[htbp]
\includegraphics[width=240pt]{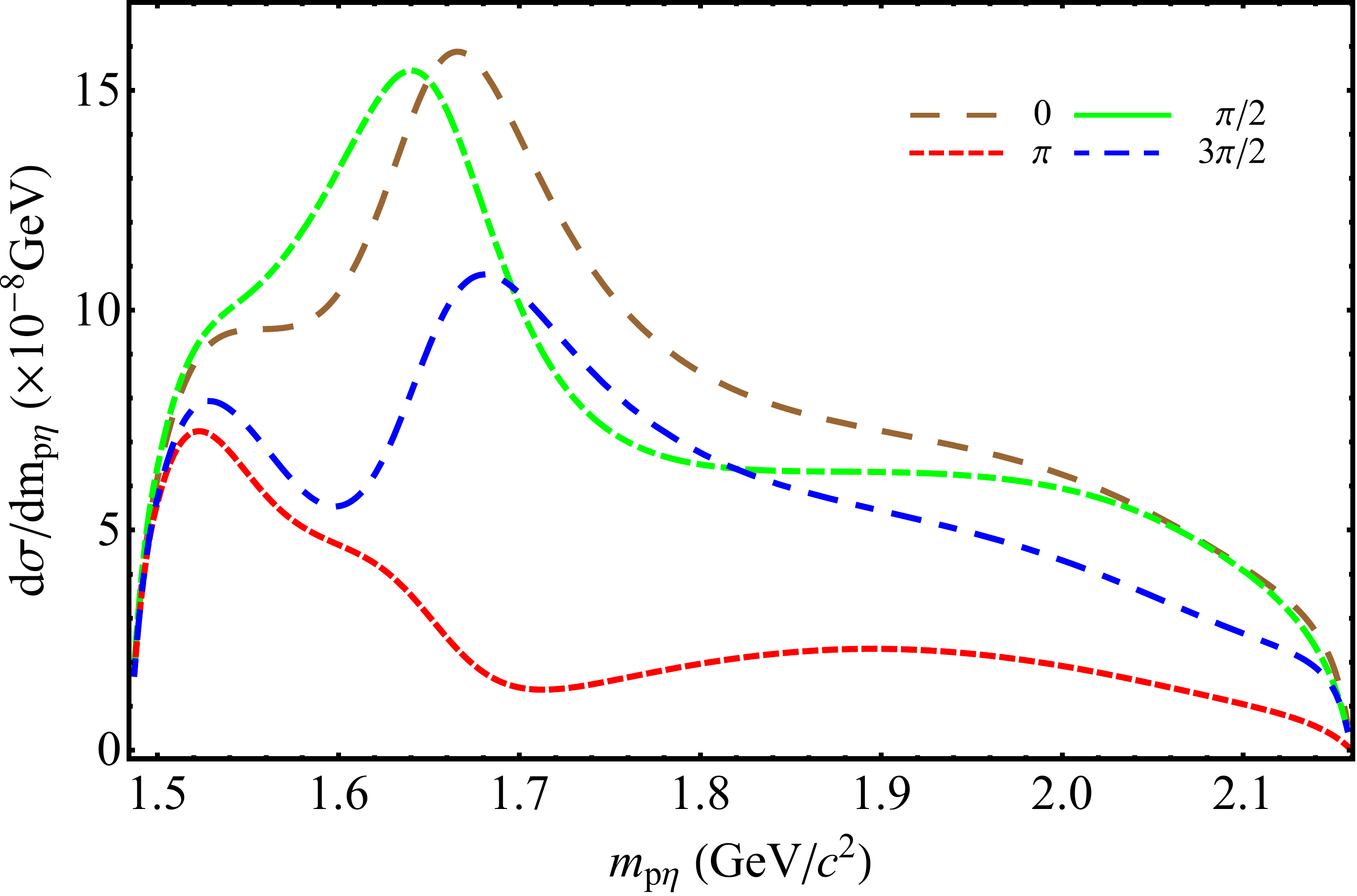}
\caption{The $p\eta$ invariant mass spectrum of the
process of $J/\psi\rightarrow p\bar{p}\eta$. The different dashed curves stand for the results obtained by considering different relative phases (0, $\pi/2$, $\pi$, $3\pi/2$) between $N(1535)$ and $N(1650)$.}\label{rephase1S}
\end{figure}

For the case of $J/\psi\rightarrow p\bar{p}\eta$ decay, we do not have available experimental data to determine the unknown $\beta$ and relative phase $\phi$ parameters. First, we calculate the invariant mass distribution of $J/\psi \to \bar{p}p \eta$ without considering the interference between different $N^*$ resonances. The numerical results obtained with $\beta=1.42$ are shown in Fig. \ref{fittingresulttot1S}, where one can see that $N(1535)$ and $N(1650)$ have significant contributions. Second, we consider only the contributions from $N(1535)$ and $N(1650)$, and we choose four typical values of $0$, $\pi/2$, $\pi$, and $3\pi/2$ for the relative phase between them. In Fig. \ref{rephase1S}, we can see that phase interference will greatly change the line shape of the $p\eta$ invariant mass distribution.
These different line shape behaviors can provide valuable information for future experimental analyses on the process $J/\psi\rightarrow \bar{p}p\eta$.

Next, we pay attention to the $p\bar{p}\rightarrow J/\psi \eta$ reaction. Although the present existing experimental data are not enough to determine the relative phases between different scattering amplitudes, we can
still accurately estimate the absolute magnitude of cross sections from different nucleon resonances.

\begin{figure}[htbp]

\includegraphics[width=240pt]{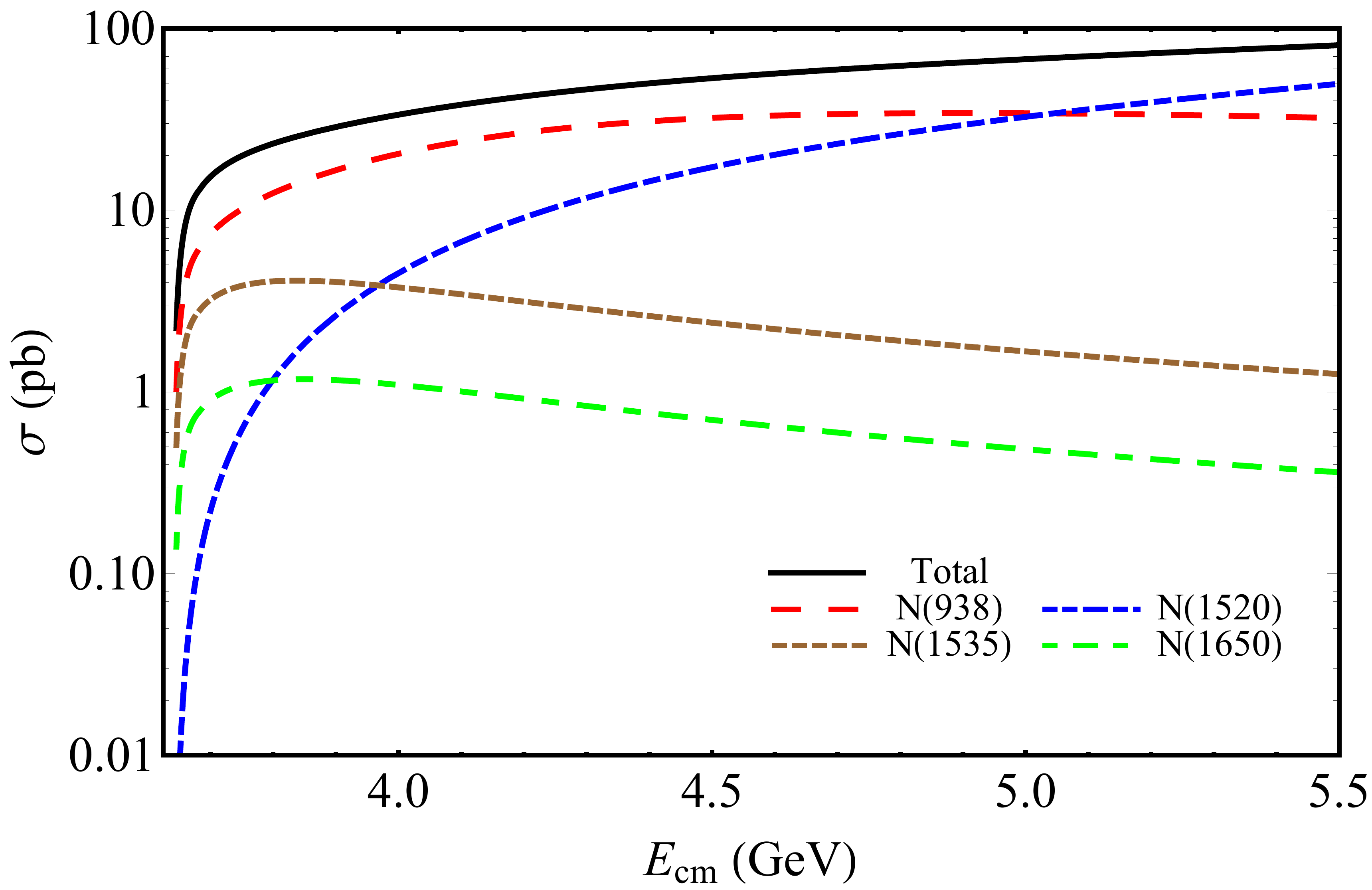}
\caption{As in Fig.~\ref{production2S}, but for the case of a $\bar{p}p \to J/\psi \eta$ reaction.}\label{production1S}.
\end{figure}

In Fig.~\ref{production1S}, we show the numerical results of total cross sections of the $p\bar{p}\rightarrow J/\psi \eta$ reaction as a function of $E_{cm}$, where the relative phases between different nucleon states are not taken into account. The results indicate that the total cross section has a maximum of about 80.5 pb at $E_{cm}=5.5$ GeV. From Fig. \ref{production1S}, one can also clearly see that contributions from excited nucleon resonances are significant, and the contribution from $N(1520)$ even exceeds the contribution of the nucleon pole when $E_{cm} > 5.0$ GeV.

\begin{figure}[htbp]
\begin{tabular}{cccc}
\includegraphics[width=115pt]{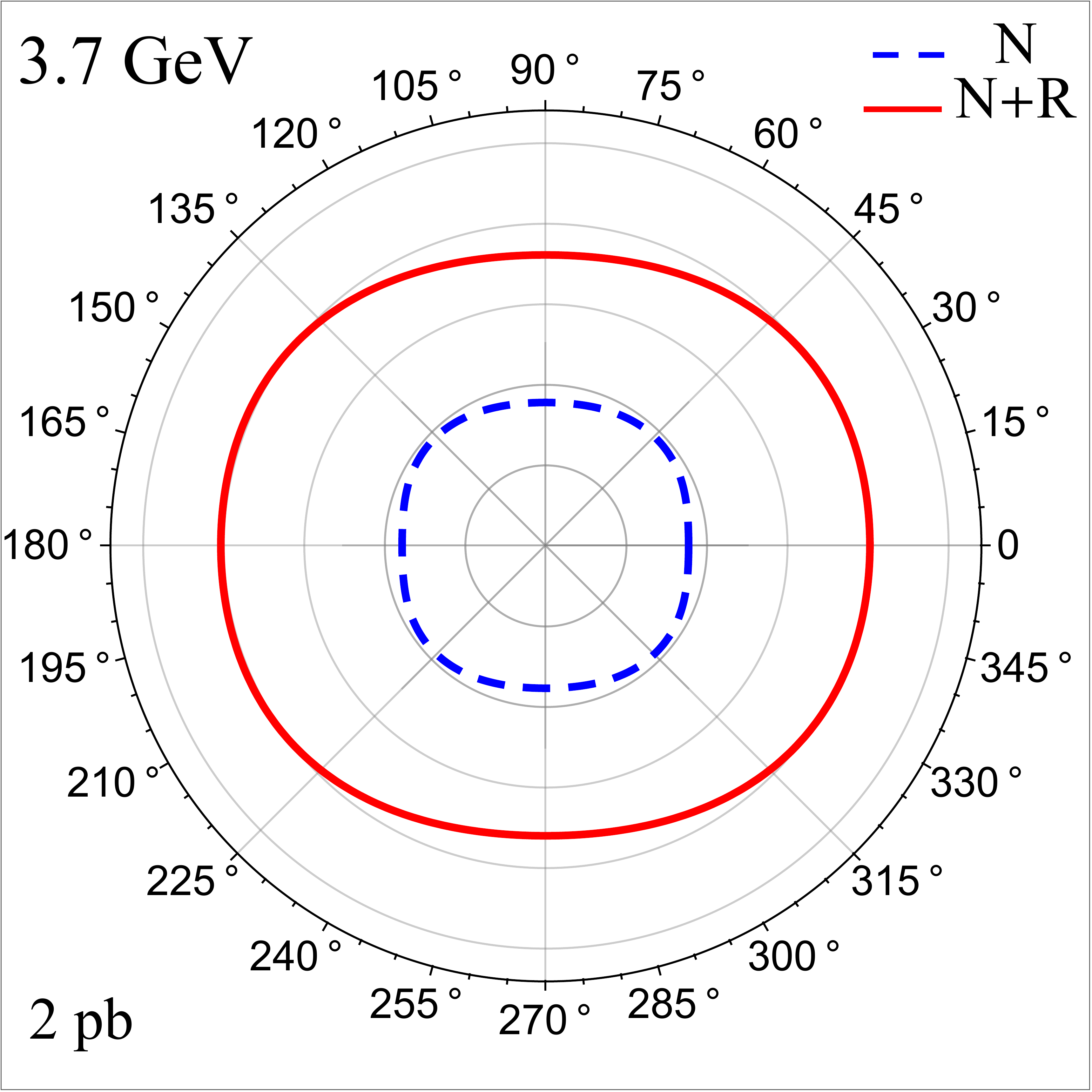}&\includegraphics[width=115pt]{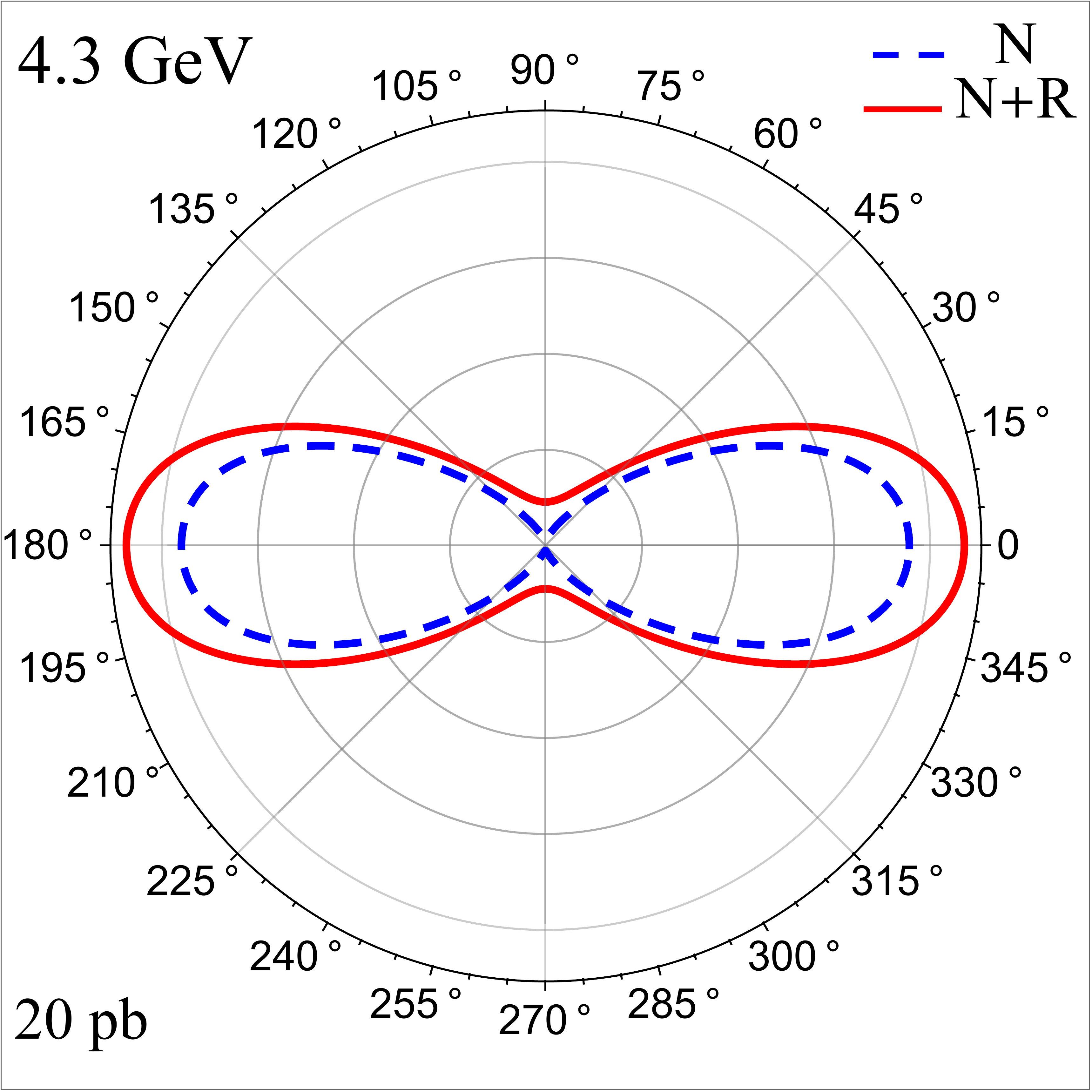}\\
\includegraphics[width=115pt]{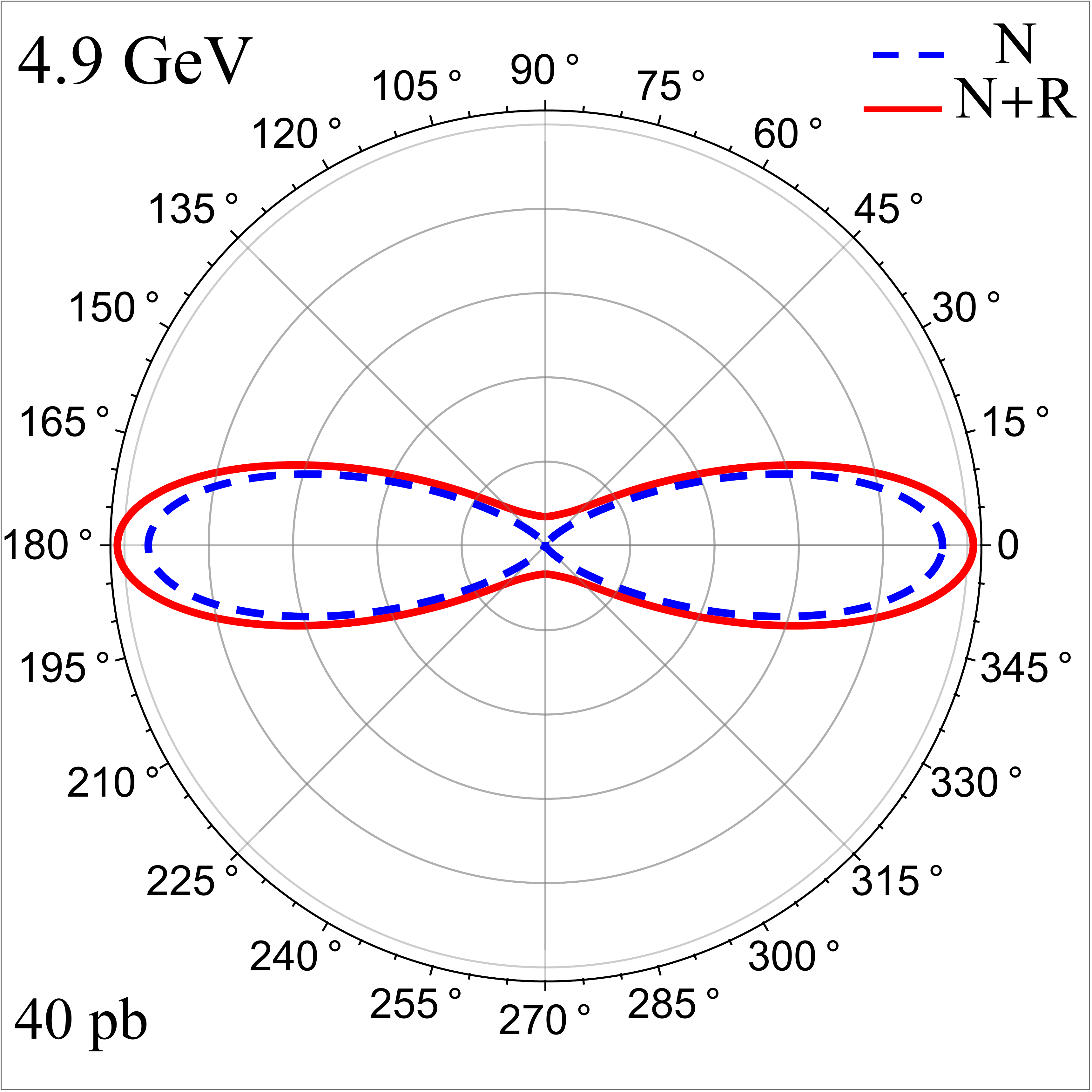}&\includegraphics[width=115pt]{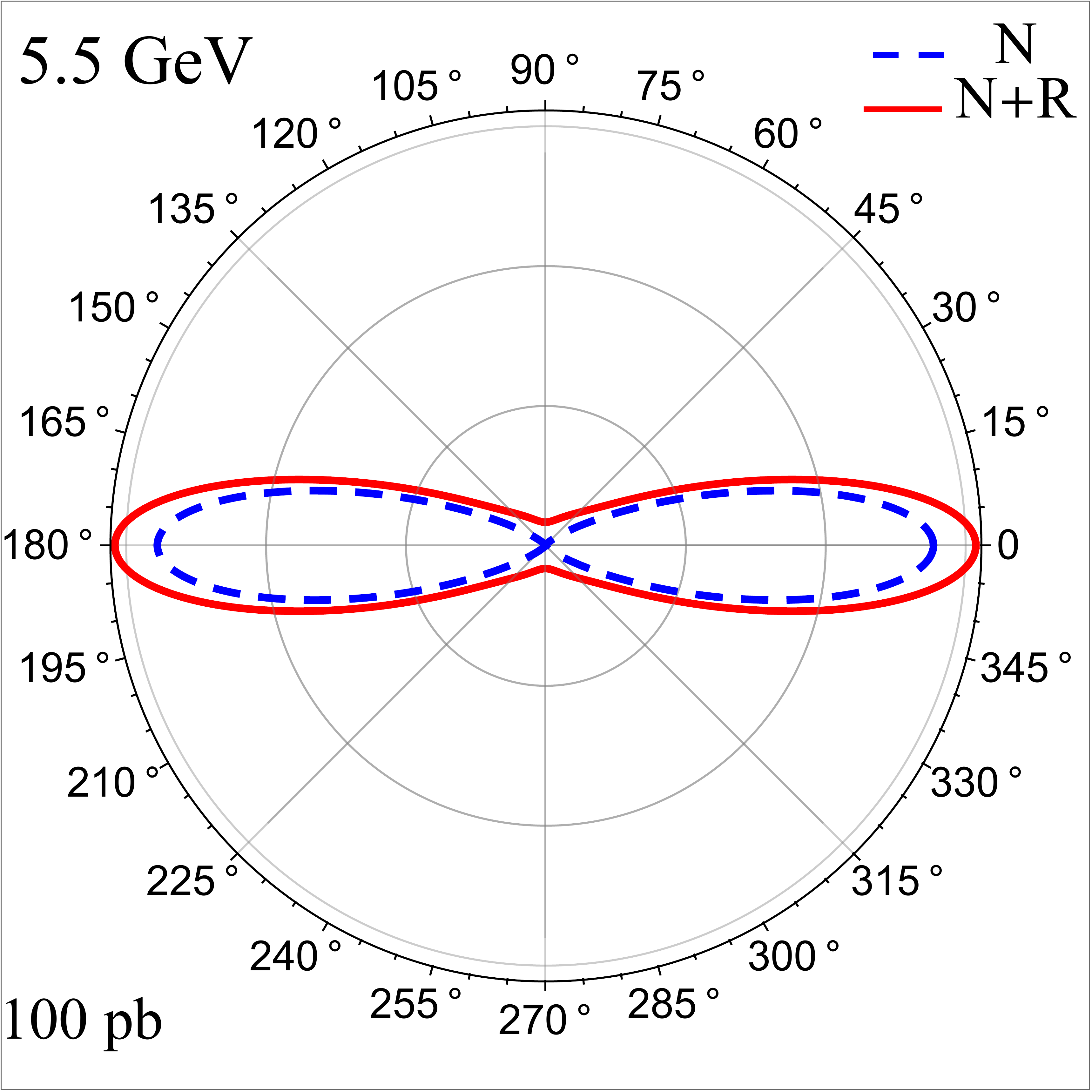}
\end{tabular}
\caption{The angular distribution of the process $p \bar{p}\rightarrow J/\psi \eta$ at different values of $E_{cm}$ (labeled in the top left corner of each panel). The red solid line is the result of considering the total contribution from the nucleon pole and nucleon resonances; the blue dashed line is the result of only considering the contribution of the nucleon pole. Each gray concentric circle denotes a specific value of $d \sigma/d \cos \theta$; from inside to outside, two neighboring concentric circles have the same value added (labeled in the bottom left corner of each panel).}\label{angulardis1S}
\end{figure}

In addition, we also calculate the angular distributions of the process $p\bar{p}\rightarrow J/\psi \eta$, which are presented in Fig. \ref{angulardis1S}.
Compared with the results for the process of $p\bar{p}\rightarrow \psi(3686) \eta$, one can see that the angular distributions of the $p\bar{p}\rightarrow J/\psi \eta$ reaction are always symmetric
with respect to $\theta=90\degree$ or $\theta=270\degree$ for all the energies that we take. This is because we did not consider the possible interference contributions among different scattering amplitudes of nucleons.
Combining this with the angular distributions of $p\bar{p}\rightarrow \psi(3686) \eta$ and $p\bar{p}\rightarrow J/\psi \eta$ reactions, we can conclude that the weight difference between the $u$ channel and $t$ channel is also due to the interference amplitudes from relative phases between different nucleon states.

\subsection{Comparison with other work}
The $p\bar{p}\rightarrow \psi \eta$ reactions are also studied in Refs.~\cite{Lundborg:2005am,Lin:2012ru}.
In Table \ref{comparison}, we make a comparison between our results and other theoretical results of Refs. ~\cite{Lundborg:2005am,Lin:2012ru}.
For the reaction of $p\bar{p}\rightarrow \psi(3686) \eta$ and $p\bar{p}\rightarrow J/\psi \eta$, our results are smaller than those of Ref. \cite{Lundborg:2005am}, but larger than the ones of Ref. \cite{Lin:2012ru}.
\begin{table}[htbp]
\centering
\caption{The total cross sections of $p\bar{p}\rightarrow \psi \eta$ estimated at $E_{cm} = 5.38$ and $4.57$ GeV in this work (the second column), given in Ref. \cite{Lundborg:2005am} (the third column) and Ref. \cite{Lin:2012ru} (the fourth column).}\label{comparison}
\begin{tabular}{ccccccccc}
\toprule[1.5pt]
\midrule[1pt]
Reaction  & This work (pb) & Ref. \cite{Lundborg:2005am} (pb) & Ref. \cite{Lin:2012ru} (pb) & $E_{cm}$\\
\midrule[1pt]
$p\bar{p}\rightarrow\psi(3686)\eta$ &  23 &  $33\pm8$ &  9 & 5.38\\
$p\bar{p}\rightarrow J/\psi\eta$ & 56& $1520 \pm 140$ & 36 & 4.57\\
\midrule[1pt]
\bottomrule[1.5pt]
\end{tabular}
\end{table}

In Ref. \cite{Lundborg:2005am}, the authors estimated the total cross sections of $p\bar{p}\rightarrow \psi X$ assuming a constant amplitude for the $\psi \rightarrow p\bar{p} X$ decay. Under this approximation, it implies that the contributions of these intermediate resonances in the decay and production process are the same. Hence, this approximation may lead to overestimation of the cross sections of $p\bar{p}\rightarrow \psi \eta$. In this work, one can clearly see that the contributions from different intermediate resonances are very different in decay and production processes, especially for $N(1535)$. The $N(1535)$ has an extremely significant contribution in the decay process of
$\psi \rightarrow p\bar{p} \eta$, but it is not important for the production process of $p\bar{p}\rightarrow \psi \eta$.
However, the constant amplitude approximation provides a good idea that the information of $p\bar{p}\rightarrow \psi \eta$ can be extracted from the decay process $\psi \rightarrow p\bar{p} \eta$.

In Ref. \cite{Lin:2012ru}, the authors first introduced the form factor in predicting the cross sections of $p\bar{p}\rightarrow \psi \eta$, where only the  nucleon pole contribution is included. Their results are shown in the fourth column of Table \ref{comparison}, and they
are smaller than our results because the contributions from nucleon resonances are considered in our numerical results.

Finally, it needs to be emphasized that we take the same form factors for both the $p\bar{p} \rightarrow \psi \eta$ reaction and the decay process $\psi \rightarrow \bar{p}p \eta$, although they should be different in these two different processes. However, since the hadron structure is still an open question, the hadronic form factors are generally adopted phenomenologically. Of course, the reliability of the treatment here can be left to future experiments to test.

\section{Summary}\label{sec5}

The forthcoming $\bar{\mbox{P}}$ANDA will be an ideal platform to carry out the study of hadron physics. Among these running facilities of particle physics, BESIII can provide abundant experimental data to the field of charm tau physics. In fact, these studies on $\bar{\mbox{P}}$ANDA and BESIII can be borrowed from each other, which was indicated in Refs. \cite{Xu:2015qqa,Wang:2017sxq}.

In this work, based on the studies of the process $\psi\rightarrow \bar{p}p\eta$, we have calculated the total cross sections and angular distributions of the $p\bar{p}\rightarrow \psi \eta$ reaction within an effective Lagrangian approach. These contributions from the nucleon resonances $N(1520)$, $N(1535)$, and $N(1650)$ in the $p\bar{p}\rightarrow \psi \eta$ reaction are considered for the first time. Our results show that these contributions from excited nucleon resonances are very important for estimating the cross section of $p\bar{p}\rightarrow \psi \eta$, and the relative phases between different amplitudes will influence the total cross section and change the shape of the angular distributions. Hence, the $\psi\rightarrow \bar{p}p\eta$ reactions are suitable for investigating the properties of the low-lying nucleon resonance.

Finally, we would like to stress that the predictions here are very
qualitative, since the contributions from ISI of $\bar{p}p$ are
neglected. We hope that these theoretical calculations presented in
this work may stimulate experimentalists' interest in exploring the
$p\bar p\to \psi \eta$ reaction through the $\bar{\mbox{P}}$ANDA
experiment. Meanwhile, we also suggest that our colleagues pay more
attentions to  the theoretical issues around the charmonium
production via the $p\bar{p}$ scattering processes, since the present
work is only a starting point.


\section*{Acknowledgments}

This work is partly supported by the National Natural Science
Foundation of China under Grants No.~ 11735003 and No. 1191101015, by
the Fundamental Research Funds for the Central Universities, and by
the Youth Innovation Promotion Association CAS (No.~2016367). X.L. is
also supported by the China National Funds for Distinguished Young
Scientists under Grant No. 11825503 and by the National Program for
Support of Top-notch Young Professionals.

\end{document}